\documentstyle[11pt,paspconf,epsf]{article}
\def\spose#1{\hbox to 0pt{#1\hss}}
\def\lap{\mathrel{\spose{\lower 3pt\hbox{$\mathchar"218$}}
     \raise 2.0pt\hbox{$\mathchar"13C$}}}
\def\gap{\mathrel{\spose{\lower 3pt\hbox{$\mathchar"218$}}
     \raise 2.0pt\hbox{$\mathchar"13E$}}}
 
\begin{document}
\title{Black Holes and Galaxy Dynamics}
\author{David Merritt}
\affil{Department of Physics and Astronomy, Rutgers University}
 
\begin{abstract}
The consequences of nuclear black holes for the 
structure and dynamics of stellar spheroids are reviewed.
Slow growth of a black hole in a pre-existing core produces a
power-law cusp, $\rho\sim r^{-\gamma}$, with $\gamma\approx 2$,
similar to the steep cusps seen in faint elliptical galaxies.
The weaker cusps in bright ellipticals, $\gamma\lap 1$, may
result from ejection of stars by a coalescing black-hole
binary; there is marginal kinematical evidence for such a process
having occurred in M87.
Stellar orbits in a triaxial nucleus are mostly regular at radii where
the gravitational force is dominated by the black hole, $r\lap r_g$;
however the orbital shapes are not conducive to reinforcing the triaxial
figure, hence nuclei are likely to be approximately axisymmetric.
In triaxial potentials, a ``zone of chaos'' extends from
a few times $r_g$ out to a radius where the enclosed stellar
mass is $\sim 10^2$ times the mass of the black hole;
in this chaotic zone, no regular, box-like orbits exist.
At larger radii, the phase space in triaxial
potentials is complex, consisting of stochastic orbits as
well as regular orbits associated with stable resonances.
Figure rotation tends to increase the degree of stochasticity.
Both test-particle integrations and $N$-body simulations
suggest that a triaxial galaxy responds globally to the presence of
a central mass concentration by evolving toward more axisymmetric shapes;
the evolution occurs rapidly when the mass of the central object
exceeds $\sim 2\%$ of the mass in stars.
The lack of significant triaxiality in most early-type
galaxies may be a consequence of orbital evolution induced
by nuclear black holes.
 
\end{abstract}
 
\section{Introduction}
 
The idea that supermassive black holes are generic components of
galactic nuclei has come to be widely accepted, due largely to
the kinematical detection of dark objects with masses
$10^{6-9.5}$M$_{\odot}$ at the centers of about a dozen galaxies
(Kormendy \& Richstone 1995; Jaffe, this volume).
The mean mass of these objects -- of order $10^{-2.5}$ times
the mass of their host spheroids -- is consistent with the
mass in black holes needed to produce the observed energy density in
quasar light given reasonable assumptions about the efficiency of
quasar energy production (Chokshi \& Turner 1992).
The black hole paradigm also explains in a natural way many of the
observed properties of energy generation in active galactic
nuclei and quasars (Blandford 1990).
However it has long been clear that supermassive black holes
might be important from a purely stellar-dynamical point of view:
both within the nucleus, where the gravitational force is
dominated by the black hole; but also at much larger radii,
if a substantial number of stars are on orbits that carry them
into the center (Gerhard \& Binney 1985).
Recent work, reviewed here, has developed these ideas and given
support to the view that supermassive black holes may be important
for understanding many of the systematic, large-scale properties
of elliptical galaxies and bulges.
 
\section{Nuclear Dynamics}
 
\subsection{Cusp formation}
 
The luminosity densities of early-type galaxies and bulges
are well approximated as power-laws,
$\nu\propto r^{-\gamma}$, inside of a ``break radius'' $r_b$
(Crane et al. 1993; Gebhardt et al. 1996).
The break radius is difficult to measure in fainter galaxies
whose steep cusps have roughly the same power-law index
($\gamma\approx 2$) as the larger-radius profile.
In bright galaxies, $M_v\lap -20$, the central cusps are shallower
($\gamma\lap 1$) and the luminosity profiles show a definite
change in slope at $r\approx r_b$ (Kormendy, this volume).
In these galaxies, $r_b$ scales roughly with total luminosity
(Faber et al. 1997);
spheroids with $M_v\approx -21$ have $r_b\approx 50$ pc,
easily resolvable from the ground for nearby galaxies.
 
To what extent are the stellar cusps attributable to the
presence of a black hole?
The gravitational force from a black hole of mass $M_h$
would dominate the force from the stars within a radius $r_g$
such that $M_*(<r_g) = M_h$.
This radius is roughly comparable to $r_b$
in the handful of galaxies for which both $r_b$ and $M_h$ can be
accurately measured;
for instance, in M87, where $M_h$ is well
constrained by the kinematics of a gas disk
(Macchetto et al. 1997), $r_g\approx r_b\approx 300$ pc.
However the black hole can strongly influence the motions
of stars only inside the (typically smaller) radius
$r_h=GM_h/\sigma_*^2$, the ``radius of influence,''
where orbital velocities around the black hole are comparable
to stellar velocity dispersions.
M87 has $r_h\approx 60$ pc, much smaller than $r_g$ or $r_b$
and barely resolvable from the ground.
 
One expects to observe photometric and kinematic features
in the stellar distribution near $r_h$; indeed, the upturn in
stellar velocities that occurs near this radius is
one signature of a black hole.
However few if any galaxies exhibit a clear
feature in the stellar luminosity profile inside of $r_b$
(Gebhardt et al. 1996),
a fact that must be explained by any model of cusp formation.
 
\begin{figure}[t]
\plotfiddle{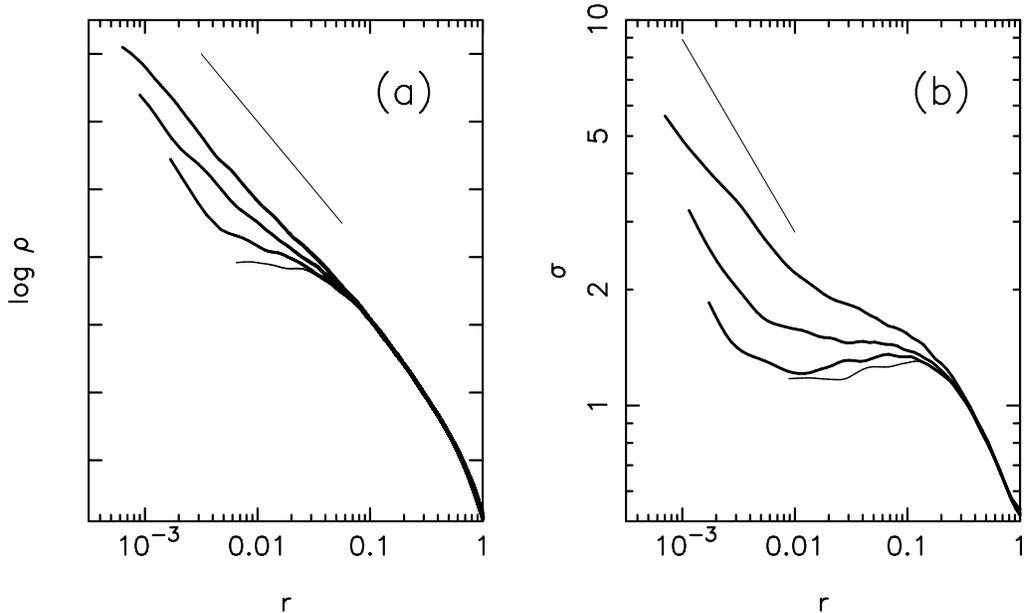}{7.75cm}{0}{75}{75}{-245}{-225}
\caption{\footnotesize
Cusp formation by adiabatic growth of a black hole
(Merritt \& Quinlan 1998).
The initial model (thin curves) is a triaxial ellipsoid,
shown here spherically symmetrized; heavy curves are
the final models after growth of a central point containing
$0.3\%$, $1\%$ and $3\%$ of the stellar mass.
{\bf (a)} Stellar density profiles $\rho(r)$;
the thin line has a logarithmic slope of $-2$.
These cusps are steeper than the $\rho\propto r^{-1.5}$
cusps that form in initially isothermal cores.
{\bf (b)} Velocity dispersion profiles $\sigma(r)$.
The small-radius dependence is $\sigma\sim r^{-1/2}$
(thin line).
}
\label{fig1}
\end{figure}
 
One widely-discussed model for the formation of stellar cusps is
the slow growth of a black hole in a pre-existing,
constant-density core.
``Slow'' here means on a time scale long compared to stellar
orbital periods, $\sim 10^{6-7}$ yr, thus guaranteeing conservation of
orbital actions; the assumption is a reasonable one if
black holes grew on the Salpeter
time scale, $t_s\approx 10^{7-8}$ yr (Krolik 1999).
If the core was initially isothermal, and if the final mass of the
black hole is less than the initial core mass, the final
density satisfies
$\rho\propto r^{-\gamma},\ \gamma=1.5$,
for $r\lap r_h$ (Peebles 1972; Young 1980).
An interesting feature of the Peebles-Young
model is the smooth, nearly
inflectionless form of the final density profile between
$r_h$ and $r_c$, the initial core radius.
Van der Marel (1999; this volume) took advantage of this fact, associating
$r_c$ with the observed break radius $r_b$.
He was able to fit the weak power-law cusps of bright ellipticals
to the region $r_h\lap r\lap r_c$ where the Peebles-Young profile is
locally well approximated by a shallow power law.
The steep cusps of faint ellipticals were less accurately
reproduced.
 
The initial conditions adopted by Peebles, Young and van der
Marel -- isothermal spheres with large core radii -- are
computationally convenient but not very compelling from a
physical point of view.
Formation of galaxies through hierarchical clustering
(e. g. Primack et al., this volume) or collapse
(e.g. May \& van Albada 1984)
tends to make systems with small or nonexistent cores and with
phase space densities
that rise more rapidly than $\sim e^{-aE}$ near the center.
Growing a black hole in such a galaxy produces a steeper cusp
than in an initially isothermal core, typically with
$\gamma\gap 2$ (Quinlan et al. 1995; Merritt \& Quinlan 1998;
Fig. 1).
This is just the slope characteristic of the central regions of
faint ellipticals and it would seem reasonable
to attribute the cusps in these galaxies to black holes.
The absence of a prominent break radius in faint ellipticals
would imply that the initial core mass (if there was a core)
did not greatly exceed the final mass of the black hole.
 
Dissipation is often invoked as a possible mechanism for
making steep cusps (e.g. Faber et al. 1997) though simulations
of dissipative core formation  (e.g. Mihos \& Hernquist 1994)
have so far failed to produce power-law profiles.
 
The weak cusps in brighter ellipticals are not so naturally
explained via the adiabatic growth model:
not only are they much shallower than
$\rho\sim r^{-2}$ but --  as Fig. 1 shows -- the transition from
$\rho\sim r^{-2}$ at $r\lap r_h$ to $\rho\sim $ constant at
$r\sim r_c$ leaves an inflection in the density profile
if $r_h < r_c$,
and such inflections are rarely if ever seen.
A more natural formation model for weak cusps would predict only one
characteristic radius, $r_b$.
Such a model was proposed by Ebisuzaki, Makino and Okumura (1991).
These authors noted that the coalescence of two black holes following
a galaxy merger would transfer energy from the binary to the stars
in the nucleus, creating a low-density core with mass comparable
to the combined mass of the two black holes.
In their model (further developed in
Makino \& Ebisuzaki 1996 and Makino 1997),
the break radius $r_b$ measures the size of the region ``scoured out''
by the binary black hole -- consistent with the observed, rough
equality of $r_b$ and $r_g$.
The predicted density profile within $r_b$ is tolerably
close to a power law with index $\gamma\lap 1$;
in this model, the observed trend of decreasing $\gamma$ with
increasing luminosity would simply reflect a greater role for
mergers in the formation of brighter galaxies.
 
One way to discriminate between the adiabatic growth and
binary black hole models for cusp formation
is via their very different
predictions about the stellar kinematics near the black hole.
Slow growth of a black hole leaves the shape of the stellar velocity
ellipsoid nearly unchanged (Young 1980; Goodman \& Binney 1984),
even at radii where $\sigma_*$ increases substantially; the reason is
that orbital eccentricities are almost unaffected by
adiabatic changes in the potential (Lynden-Bell 1963).
By contrast, ejection of stars by a coalescing black hole binary
produces a core with strongly circularly-biased motions, since
stars on radial trajectories are more likely to interact with the
binary (Quinlan 1996).
Quinlan \& Hernquist (1997) and Nakano \& Makino (1999)
found that the velocity anisotropy
$\beta \equiv 1-\sigma_t^2/\sigma_r^2$ in an initially
isotropic core can drop as low as $\sim -1$ after the ejection
of stars is complete.
 
\begin{figure}
\plotfiddle{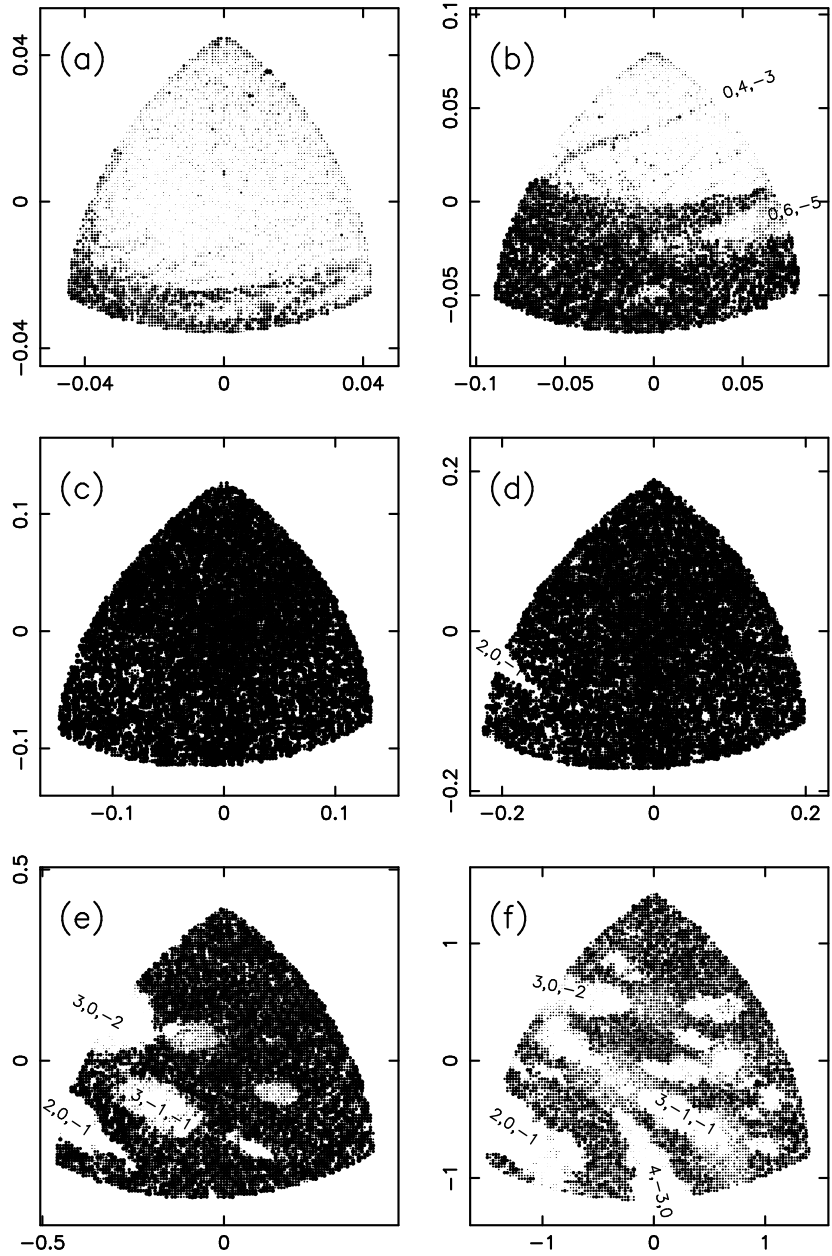}{16.5cm}{0}{75}{75}{-215}{-80}
\caption{\footnotesize
Properties of box-like orbits in a triaxial potential
containing a central point mass.
Each panel shows one octant of an equipotential surface;
the $z$ (short) axis is vertical and the $x$ (long) axis is to
the left.
Orbits were started on this surface with zero velocity.
The degree of stochasticity is indicated by the grey scale;
white regions correspond to regular orbits.
Resonance zones are labelled by their defining integers.
The ratio $M_h/M_*$ of black hole mass to enclosed stellar mass is
{\bf (a)} 0.33, {\bf (b)} 0.22, {\bf (c)} 0.13, {\bf (d)} 0.088, 
{\bf (e)} 0.032 and {\bf (f)} 0.0079.
The half-mass radius of the model is approximately one.
Near the black hole (a), orbits are mostly regular;
in the ``zone of chaos'' (c, d), almost all box-like 
orbits are chaotic;
and at large radii (f), regular and stochastic orbits
co-exist.
}
\label{fig2}
\end{figure}
 
Stellar velocity anisotropies are difficult to extract from
integrated spectra; the task becomes much easier
if the form of the gravitational potential is known a priori,
since the anisotropy then follows directly from the line-of-sight
velocity dispersion profile (Binney \& Mamon 1982).
The velocity polarization predicted by the binary black hole model
is therefore best looked for in a galaxy where $M_h$
has been determined independently from the stellar data.
M87 is such a galaxy, and in fact the ground-based stellar
data, combined with the Macchetto et al. (1997) estimate of $M_h$,
imply a substantial anisotropy, $\beta\approx -1$, at $r\lap r_h$
(Merritt \& Oh 1997).
However the statistical significance of the result is small due
to the low central surface brightness of this galaxy.
Planned observations of M87 and other galaxies with STIS on HST
should soon resolve this issue.
 
\subsection{Nonspherical nuclei}
 
The gravitational potential in the vicinity of the black hole, 
at $r\lap r_g$, is nearly Keplerian; 
forces from the stars in the cusp constitute a small perturbation, 
causing orbits to precess.
In an axisymmetric galaxy, this precession converts an otherwise
closed, elliptical orbit into a tube orbit which fills a
doughnut-shaped region around the symmetry axis.
Tube orbits avoid the center due to conservation of angular
momentum;
hence a star on a tube orbit can come only so close to the black hole.
The situation is very different, and much more interesting, in
non-axisymmetric or triaxial nuclei.
While tube orbits still exist in such potentials,
orbits that pass arbitrarily close to the center
are possible as well.
 
Figure 2 illustrates how the character of ``box-like'' orbits --
orbits with stationary points and (in an integrable potential) 
filled centers -- varies with distance from the
central black hole in a triaxial potential.
Near the black hole, $r\lap r_g$, almost all box-like
orbits are regular, i.e. non-chaotic, respecting three isolating integrals
of the motion.
Such orbits in the planar geometry have been dubbed ``lenses''
by Sridhar \& Touma (1999);
in three dimensions, the orbits are shaped like
pyramids with rectangular bases.
The black hole lies just
inside the vertex of the pyramid, at the focus of the precessing
ellipse, and the pyramid's central axis is coincident with the short ($z$)
axis of the triaxial figure.
A superposition of two such orbits, oriented above and below the
$x-y$ plane, is symmetric and looks very much like the regular
box orbits in integrable triaxial potentials (de Zeeuw 1985).
However the latter are aligned with the {\it long} axis of the triaxial
figure, making them much more useful 
for self-consistently reconstructing the stellar density.
 
At larger radii, $r\gap r_g$, the potential is no longer approximately
Keplerian and the integrability is lost -- box-like orbits are
generically stochastic (Fig. 2c, d).
This ``zone of chaos'' in triaxial potentials extends from a few times
$r_g$ outward to much larger radii, as discussed below.
 
Could the predominantly regular orbits -- pyramids and tubes --
within $\sim r_g$ be used to self-consistently
reconstruct a triaxial nucleus containing a central black hole?
The question is in principle straightforward to answer
though as yet no attempts have been made.
Two studies (Kuijken 1993; Syer \& Zhao 1998) addressed the
self-consistency problem for scale-free, non-axisymmetric disks
with divergent central densities.
The major families of box-like orbits in both of these
planar potentials are $2:1$ ``bananas'';
in spite of having a favorable orientiation parallel to the long
axis, the banana orbits were found to be too limited in
their range of shapes to reproduce the assumed figure.
Pyramid orbits have even less favorable
shapes and it is reasonable to expect the triaxial
self-consistency problem for black hole nuclei to be
at least as narrowly constrained as that for scale-free disks.
 
It therefore seems likely that significant triaxiality
would be difficult to maintain near the center of a galaxy
containing a supermassive black hole: both at radii $r\lap r_g$,
because of the unfavorable orientation of the (predominantly
regular) orbits; and at radii $r\gap r_g$, because of chaos.
 
In advance of more definite theoretical predictions,
a number of workers have recently investigated the
isophotal shapes of early-type galaxies at radii near $r_b$.
Quillen (this volume) finds a change in ellipticity and boxiness
between $\sim r_b$ and a few times $r_b$ in the isophotes of two
galaxies; the change is in the direction of more
elongated and boxier isophotes at large radii.
Tymann (cited in Bender, this volume) also finds less boxy
isophotes inward of $r_b$ in a sample of early-type galaxies.
Ryden (1998; this volume) shows that bright ellipticals as a
class exhibit rounder isophotes at radii of a few times $r_b$
than at larger radii;
the effect is consistent with more nearly axisymmetric shapes at
smaller radii, since oblate spheroids are more
likely than triaxial ellipsoids to appear round under random
projection.
 
These results suggest a possible change in the shapes or
orbital compositions of galaxies near $r_b$, perhaps in
the direction of more axisymmetric configurations at smaller
radii.
As Quillen and Ryden both note, such changes could reflect
the constraints that a black hole imposes on the shapes of
orbits, or they could be relics of the core formation process,
or both.
Distinguishing between these possibilities will be easier once
the self-consistency problem for black-hole nuclei is better
understood.
 
Orbits like the pyramids are not centered on the black
hole, a consequence of the near-Keplerian nature of the potential.
Sridhar \& Touma (1999) noted that off-center orbits can persist even
in nuclei where the black hole itself is offset from the center
of the stellar spheroid; furthermore they identified one orbit
family for which the orbital offset was in the same direction as that of
the spheroid.
One could imagine using such orbits to construct self-consistent,
lopsided nuclei along the lines of the conceptual model proposed
by Tremaine for M31 (1995).
One attempt, using an $N$-body code, is described by
Jacobs \& Sellwood (this volume); these authors
were unable to construct long-lived lopsided disks unless the
disk mass was less than a few percent of the mass of the black hole,
substantially smaller than the estimated mass of the stellar disk in M31
(Kormendy \& Bender 1999).
 
\section{Large-Scale Dynamics}
 
\subsection{Regular and stochastic orbits}
 
The gravitational influence of a nuclear black hole can extend
far beyond $r_g$ in a non-axisymmetric galaxy, since orbital
angular momenta are not conserved and stars with arbitrarily
large energies can pass close to the center (Gerhard \& Binney
1985).
In a triaxial potential containing a central point mass,
the phase space divides naturally into three regions depending on
distance from the center (Fig. 2).
In the innermost region, $r\lap r_g$, the potential is dominated by the
black hole and the motion is essentially regular, as discussed above.
At intermediate radii (Fig. 2c, d),
the black hole acts as a scattering center, rendering
almost all of the center-filling orbits stochastic.
This ``zone of chaos'' extends outward from a few
times $r_g$ to a radius where the enclosed stellar mass is
roughly $10^2$ times the mass of the black hole.
If $M_h$ exceeds $\sim 10^{-2}M_{sph}$,
as it appears to do in a few galaxies,
the ``zone of chaos'' includes essentially the entire potential
outside of $\sim r_g$.
However if $M_h\lap 10^{-2}M_{sph}$, there is a third, outer region
in which the phase space is a complex mixture of chaotic and regular
trajectories (Fig. 2e, f).
In the absence of a central point mass, the orbital structure of
a triaxial potential resembles this complex outer region at all
energies (Schwarzschild 1993; Merritt \& Fridman 1996;
Carpintero \& Aguilar 1998; Papaphillipou \& Laskar 1998;
Valluri \& Merritt 1998; Wachlin \& Ferraz-Mello 1998).
 
\begin{figure}[t]
\plotfiddle{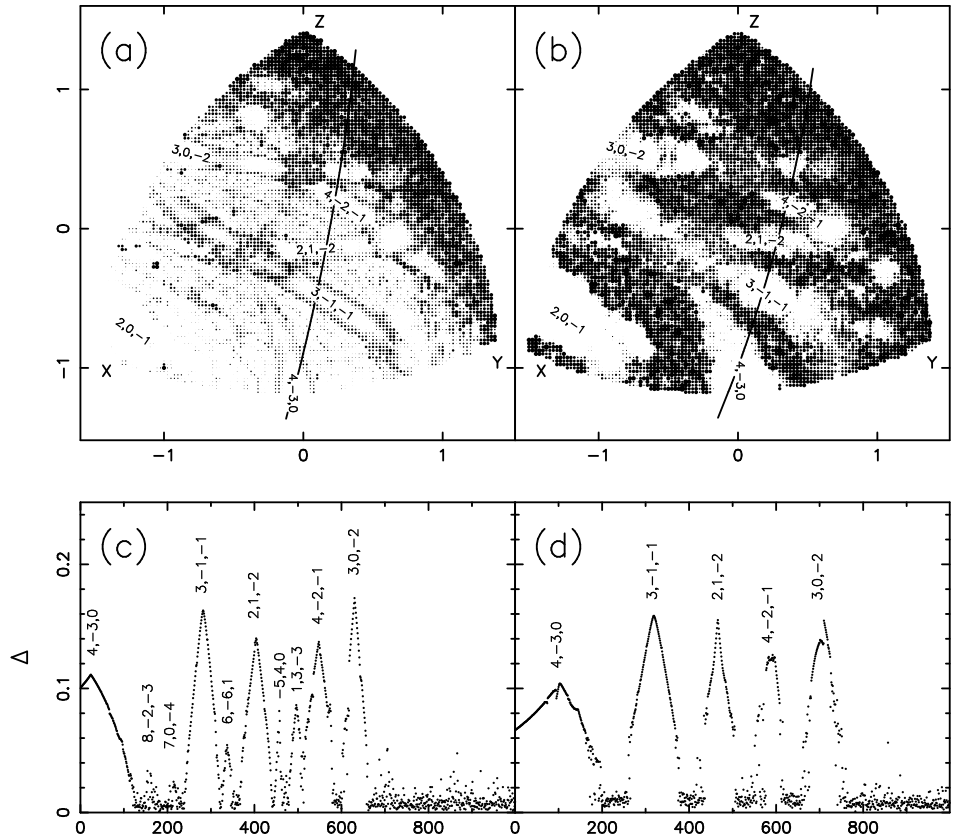}{11.cm}{0}{65}{65}{-215}{-170}
\caption{\footnotesize
Resonances in triaxial potentials (Merritt \& Valluri 1999).
The mass model in {\bf (a)} has a weak ($\gamma=0.5$) cusp and no black
hole;
in {\bf (b)} the black hole contains $0.3\%$ of the total mass.
Both equipotential surfaces lie close to the half-mass radius.
The grey scale measures the degree of stochasticity of orbits
started with zero velocity on the equipotential surface,
as in Fig. 2.
Stable resonance zones -- the white bands in (a) and (b) -- are
labelled by the order $(m_1,m_2,m_3)$ of the resonance.
Panels {\bf (c)} and {\bf (d)} show the pericenter distance 
$\Delta$ of a set of $10^3$ orbits with
starting points along the heavy solid lines in (a) and (b).
}
\label{fig3}
\end{figure}
 
The complexity of the phase space far from the black hole
is a consequence of resonances.
A resonant orbit is one for which the fundamental frequencies
$\omega_i, {i=1,2,3}$ on the invariant torus are ``commensurate,''
satisfying a relation of the form
$m_1\omega_1 + m_2\omega_2 + m_3\omega_3 = 0$ with integer $m_i$.
In the case of 2D motion, a resonant orbit is closed,
returning to its starting point after a time
$T=2\pi |m_2|/\omega_1=2\pi |m_1|/\omega_2$.
In three dimensions, resonances do not imply closure;
instead, a resonant trajectory is confined for all time to a
two-dimensional sub-manifold of its 3-torus
(Valluri \& Merritt, this volume).
Such an orbit is thin, densely filling a membrane in
configuration space.
 
Resonant tori play roughly the same role, in three dimensions,
that closed orbits play in two, generating families of regular
orbits when stable and stochastic orbits when unstable
(Merritt \& Valluri 1999).
In triaxial potentials,
the main source of instability is gravitational deflections
from the central point mass;
stable orbits are ones that avoid the center.
Tube orbits achieve this via a $1:1$ resonance in one of the
principal planes; box orbits are generically center-filling,
but a box orbit associated with a sufficiently low-order
resonance can also avoid the center by a wide enough
margin to remain stable.
 
The degree to which this is possible depends on the
steepness of the central force gradient (Valluri \& Merritt 1998).
When the central singularity is weak -- for instance, a
$\rho\propto r^{-\gamma}$ stellar cusp -- box-like orbits can
remain stable even when their pericenter distances are small.
Phase space then consists of a large number of
intersecting and overlapping resonance zones, some of high order,
corresponding to thin orbits with many sheets (Fig. 3a).
When the central singularity is stronger -- e.g. a
central point mass -- only a handful of low-order resonances
can maintain sufficient pericenter distance to remain stable (Fig.
3b);
high-order resonances typically generate stochastic zones.
As the mass of the central point is increased, fewer and fewer of
the resonant orbits are able to avoid the center by a wide enough
margin
and the phase space undergoes a transition to global stochasticity
-- essentially all of the box-like trajectories are chaotic.
While this transition has only been studied in a handful of
model potentials,
it seems to occur whenever the black hole mass exceeds
$\sim 2-3\%$ of the enclosed mass in stars
(Merritt \& Quinlan 1998; Valluri \& Merritt 1998; Merritt \& Valluri 1999).
 
The influence of figure rotation on the orbital composition of
triaxial potentials has not yet been systematically studied.
Valluri (this volume) finds that figure rotation tends to
increase the degree of orbital stochasticity, apparently because
the Coriolis forces broaden orbits that would otherwise
be thin, driving them into the destabilizing center.
 
\subsection{Black-hole-induced evolution}
 
Stochastic motion introduces a new time scale into galactic dynamics,
the mixing time
(Kandrup \& Mahon 1994; Kandrup, this volume).
Mixing is the process by which a non-uniform distribution of
particles in phase space relaxes to a uniform distribution, at
least in a coarse-grained sense.
A weak sort of mixing, called phase mixing, occurs even in
integrable potentials, as particles on adjacent tori gradually
move apart (Lynden-Bell 1967);
phase mixing is responsible for the fact that the coarse-grained
phase space density in relaxed systems is nearly constant around
tori.
Mixing in chaotic systems can be much more effective than phase
mixing,
since stochastic trajectories are exponentially unstable
and not confined to tori.
Chaotic mixing is also irreversible in the sense that an
infinitely fine tuning of velocities would be required in order
to undo its effects.
 
Mixing driven by a central black hole converts all of the
stochastic trajectories at a single energy into an invariant
ensemble whose shape is similar to that of an equipotential
surface, hence rounder than the figure.
Two consequences are likely: the galaxy should become rounder,
or at least more axisymmetric, due to the loss of the regular
orbits needed to maintain triaxiality;
and sharp features in the phase-space distribution should be
smoothed out.
Mixing induced by a central singularity ceases if the
stellar distribution reaches an axisymmetric state since
few stars are then able to approach the destabilizing center.
 
\begin{figure}[t]
\plotfiddle{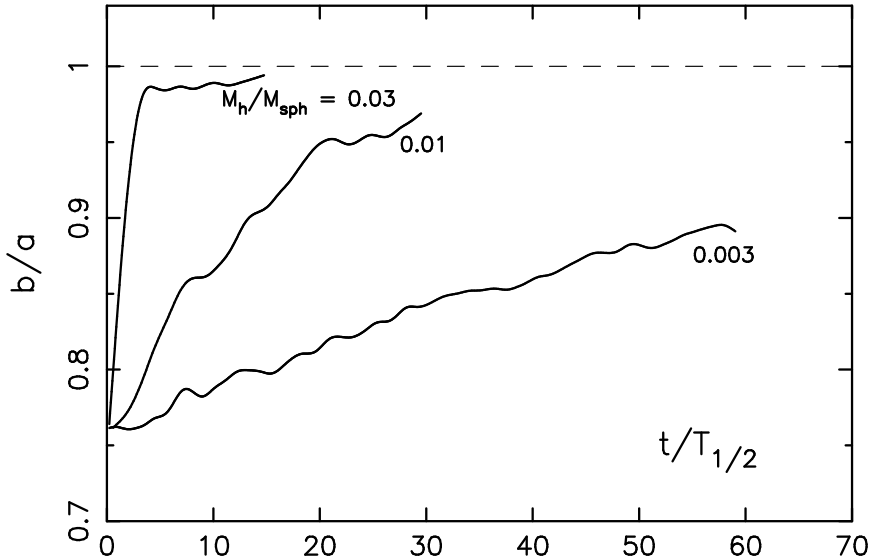}{7.5cm}{0}{60}{60}{-190}{-180}
\caption{\footnotesize
Response of an initially triaxial galaxy to growth
of a central point mass (adapted from Merritt \& Quinlan 1998).
The intermediate-to-long axis ratio $b/a$,
defined by the most-bound 50\% of the stars,
is plotted as a function of time;
$T_{1/2}$ is the period of a circular orbit at the half-mass radius
in a spherical model with the same radial distribution of mass.
The growth time of the central mass was $\sim 5T_{1/2}$ for the
two smaller values of $M_h$ and $\sim 2T_{1/2}$ for the larger value.
}
\label{fig4}
\end{figure}
 
The rate at which mixing would induce such changes can be
estimated by integrating trajectories in fixed triaxial
potentials.
Such experiments
(Kandrup \& Mahon 1994; Mahon et al. 1995;
Merritt \& Valluri 1996; Valluri \& Merritt 1998)
reveal a strong dependence of the mixing rate on the
structure of phase space.
In regions containing both regular and stochastic trajectories
-- e.g. the two outermost shells of Figure 2 -- mixing is inefficient,
presumably because the invariant tori of the regular orbits
hinder the diffusion of the stochastic orbits.
In regions where the motion is almost fully chaotic -- e.g. the
``zone of chaos'' in Figure 2 -- mixing occurs very rapidly,
in a few crossing times.
Orbits in such regions lose all memory of their
initial conditions after just a few oscillations.
 
Norman, May \& van Albada (1985) made one of the first
attempts to simulate the large-scale response
of a galaxy to the orbital evolution induced
by a central black hole.
These authors observed only a slight response at
the centers of their $N$-body models;
however their initial conditions were almost precisely
axisymmetric, thus guaranteeing that the influence of the black
hole would be limited to $r\lap r_g$.
More dramatic evolution was seen in a number of subsequent
studies of dissipative galaxy formation
(e.g. Katz \& Gunn 1991; Udry 1993; Dubinski 1994).
These authors used $N$-body codes to simulate the accumulation of
mass at the centers of initially triaxial galaxies or halos; in
each case, evolution toward more axisymmetric shapes was observed
when the central mass exceeded a few percent of the mass in
stars.
Barnes (1996; this volume) observed a similar response
in $N$-body simulations of mergers between disk galaxies:
purely stellar-dynamical mergers produced strongly triaxial
remnants,
but adding as little as $1\%$ of the mass
in the form of a dissipative component resulted in nearly
axisymmetric final shapes.
The evolution toward axisymmetry seen in these simulations is
sometimes loosely attributed to ``dissipation,'' but in fact
it is purely a stellar dynamical effect:
the stars respond to the ``gas'' only insofar as the latter
affects the gravitational potential.
 
Merritt \& Quinlan (1998) repeated the Norman et al. (1985)
experiments, using initial models that were significantly
triaxial at all radii.
They observed a global response toward axisymmetry
as the mass of the central point was increased;
the rate of evolution was found to depend strongly on the
ratio of black hole mass to galaxy mass (Fig. 4).
When $M_h/M_{sph}$ was $0.3\%$, the galaxy
evolved in shape over $\sim 10^2$ orbital periods, whereas
increasing $M_h/M_{sph}$ to $3\%$ caused the galaxy
to become almost precisely axisymmetric in little more than a crossing time.
Rapid evolution toward axisymmetry
occurred at any radius whenever the ``black hole'' mass exceeded $\sim
0.025$ times the enclosed stellar mass -- roughly the same mass
ratio at which the regular box-like orbits disappear (Fig. 2).
 
These experiments provide a natural explanation for the
absence of significant triaxiality in most elliptical galaxies
(Franx, de Zeeuw \& Illingworth 1991).
Based on Fig. 4, a galaxy with
a ``typical'' black hole, $M_h/M_{sph}\approx 0.003$, would
evolve to axisymmetry in roughly 100 periods of the half-mass
circular orbit; this time span is of order a galaxy lifetime for
elliptical galaxies with $M_v\approx -19$ or $-20$.
Fainter ellipticals have generally shorter crossing times and
hence should be weakly triaxial at best;
brighter ellipticals might still retain their
(merger-induced?) triaxial shapes.
These predictions are consistent with what little
is known about the statistics of elliptical galaxy
intrinsic shapes
(Ryden 1996; Tremblay \& Merritt 1996; Bak \& Statler 1999).
 
Orbital evolution induced by a black hole should smooth out the
stellar phase-space distribution at the same time
that it destroys triaxiality.
In fact Merritt \& Quinlan (1998) noted a striking change in the
isophotal shapes of their $N$-body models, from strongly
peanut-shaped at the start to nearly elliptical after the
black hole was in place.
Boxy or peanut-shaped isophotes are a natural consequence of
a non-smooth phase space density (Binney \& Petrou 1982).
One might therefore predict a correlation between
triaxiality and boxiness in real galaxies,
since the orbital evolution induced by a nuclear black hole
would tend to eliminate the two in tandem.
Kormendy \& Bender (1996) noted just such a correlation;
furthermore the majority of boxy ellipticals are bright
(Bender, this volume),
consistent with the expected, longer time scales for orbital
evolution in brighter galaxies.
 
It is remarkable that the minimum black hole mass required to induce
rapid evolution in the orbital composition of a triaxial ellipsoid --
$M_h/M_{sph} \approx 2\%$ -- is essentially equal to the
maximum value of $M_h/M_{sph}$ observed in real galaxies
(Kormendy et al. 1996; Cretton \& van den Bosch 1999).
This agreement could be fortuitous, or it could point to a
connection between the fueling of black holes and the shapes
of their host spheroids.

\end{document}